\documentclass[prd,
twocolumn,
superscriptaddress,nofootinbib,%
tightenlines,showpacs,showkeys]{revtex4}

\usepackage{amsfonts,amsmath,graphicx,epsfig,amssymb}
\usepackage{mathrsfs}
\usepackage{latexsym}
\usepackage{amsxtra}
\usepackage{amsbsy}
\usepackage{amscd}
\usepackage{color}
\usepackage{bbm}
\usepackage{hyperref}
\usepackage{multirow}

\usepackage{graphics}
\usepackage{amsfonts}

\usepackage{subfigure}
\usepackage{color}
\allowdisplaybreaks

\newcommand{\rbox}{\rule[-0.20cm]{0cm}{5mm}}
\newcommand{\be}{\begin{equation}}
\newcommand{\ee}{\end{equation}}

\def\no{\nonumber}
\def\bea{\arraycolsep .1em \begin{eqnarray}}
\def\eea{\end{eqnarray}}

\begin{document}
\title{\Large Test of semi-local duality in a large {\boldmath$N_C$} framework}
\author{Ling-Yun Dai}
\email{l.dai@fz-juelich.de}
\affiliation{School of Physics and Electronics, Hunan University, Changsha 410082, China}
\affiliation{Institute for Advanced Simulation, Institut f\"ur Kernphysik
   and J\"ulich Center for Hadron Physics, Forschungszentrum J\"ulich, D-52425 J\"ulich, Germany}
\author{Xian-Wei Kang}
\email{kangxianwei1@gmail.com}
\affiliation{College of Nuclear Science and Technology, Beijing Normal University, Beijing 100875, China}
\author{Ulf-G. Mei{\ss}ner }
\email{meissner@hiskp.uni-bonn.de}
\affiliation{Helmholtz Institut f\"ur Strahlen- und Kernphysik and Bethe Center
 for Theoretical Physics, Universit\"at Bonn, D-53115 Bonn, Germany}
\affiliation{Institute for Advanced Simulation, Institut f\"ur Kernphysik
   and J\"ulich Center for Hadron Physics, Forschungszentrum J\"ulich, D-52425 J\"ulich, Germany}
\begin{abstract}
In this paper we test the semi-local duality based on the method of Ref.~\cite{Dai:2017uao}
for calculating final-state interactions at varying  number of colors ($N_C$).
We compute the amplitudes by dispersion relations that respect analyticity and coupled
channel unitarity, as well as accurately describing experiment.
The $N_C$ dependence of the $\pi\pi\to\pi\pi$ scattering amplitudes is obtained by comparing
these amplitudes to the one of chiral perturbation theory.
The semi-local duality is investigated by varying $N_C$.  Our results show that the
semi-local duality is not violated when $N_C$ is large. At large $N_C$, the contributions of
the $f_2(1270)$, the $f_0(980)$ and the $f_0(1370)$ cancel that of the $\rho(770)$ in the finite
energy sum rules, while the $f_0(500)$ has almost no effect. This gives further credit to the
method developed in Ref.~\cite{Dai:2017uao} for investigating  the $N_C$ dependence of
hadron-hadron scattering with final-state interactions. This study is also helpful to understand
the structure of the scalar mesons.
\end{abstract}

\maketitle

\parskip=2mm
\baselineskip=3.5mm

\section{Introduction}\label{sec:introduction}
The $1/N_C$ expansion \cite{'tHooft:1973jz,'tHooft:1974hx} provides an effective diagnostics to
differentiate the ordinary from the non-ordinary quark-antiquark structure of the mysterious scalars,
see e.g.~\cite{Pelaez04,Pelaez06,MRP11,Sun,DLY11}.
In the physical world, i.e. at  $N_C=3$, there should be local duality
\cite{Veneziano:1968yb,Dolen:1967zz,Dolen:1967jr,Schmid:1968zza,Schmid:1968zz,Shiga:1971by}.
This means that  Regge exchange in the crossed channel is dual to the contribution of
resonances in the direct channel. One thus only needs to add either the Regge term or the direct
channel resonances in a given calculation. An explicit model shows that there is no interference
between these two contributions \cite{Veneziano:1968yb}. Indeed, in the high-energy region the
overlap of the resonances is much stronger, leading to a smooth amplitude. Such a smooth amplitude
is similar to the one generated by Regge poles in the $t$-channel.
The cross section is therefore more readily described by the Regge exchanges in the crossed
channel rather than by lots of resonances in the direct ($s-$)channel. However, in the real world, things are
more complicated as the widths of the resonances are finite and only semi-local duality is
fulfilled \cite{MRP11}. Through finite-energy sum rules (FESR) the equivalence between the
resonances in the direct $s$-channel and the Regge poles in the crossed $t$-channel
holds on the average \cite{Schmid:1968zz,Schmid:1968zza}.

In the pioneering work of \cite{MRP11}, the semi-local duality is
tested in the large $N_C$ case and it is shown to be useful for investigating the structure of
the light scalar mesons. The scattering amplitudes are
obtained by unitarized chiral perturbation theory (U$\chi$PT) and
the $N_C$ dependence of  the pertinent  low-energy constants is
taken over to the amplitudes. The FESR are tested by tuning $N_C$ up to
30 or 100. They found that the $f_0(500)$ (often also called the $\sigma$) should
contain a sub-dominant $\bar{q}q$ component and this ensures the semi-local duality to be
fulfilled up to $N_C=15-30$. This was later used to constrain the
meson-meson scattering amplitudes calculated within $U(3)$ unitary $\chi$PT~\cite{Guo:2012yt}.
The semi-local duality could be fulfilled very well up to $N_C=30$. The relation between local duality and exotic states is also
discussed in Ref.~\cite{Zheng:2004xw}.

On the other hand, final-state interactions (FSI)  play an
important role in hadron phenomenology, especially
when the energy is not very far away from the threshold of a
pair or triplet of hadrons. For different models to describe the FSI, see
e.g.~\cite{AMP-FSI,Meissner:1990kz,Guo2011,Dai:2012pb,Kang:2013uia,Kang:2013jaa,Garcia-Recio:2013uva,Chen2015,Wilson2016,Colangelo:2016jmc,Hanhart:2016pcd,Dai:2017ont,Cheng:2017pcq}.
In our earlier paper, a new method to study the large $N_C$ behavior
of the FSI was proposed \cite{Dai:2017uao}. The $N_C$ dependence is
generated based on the fact that the tangent of the phase is
proportional to $1/N_C$, that is, $\tan
\varphi\sim\mathcal{O}(1/N_C)$, where ${\rm Re}T\sim\mathcal{O}(1/N_C)$
and ${\rm Im}T\sim\mathcal{O}(1/N_C^2)$ are naturally given by
chiral perturbation theory ($\chi$PT). The trajectories of the
widths of the $\rho$ and the $f_2$ quantitatively behave as $1/N_C$, which
confirms the reliability of the method. Following that work, a
natural extension is to check whether the semi-local duality is
satisfied using this method.

This paper is organized as follows:
In Sect.~\ref{sec:T} we use a dispersive method to obtain the $I=2$ $\pi\pi$ scattering
partial waves up to $s\sim4$~GeV$^2$, that were not considered in~\cite{Dai:2017uao}.
The amplitudes are constructed analytically and respect the coupled channel unitarity
and give a good description of the experimental data.
In Sect.~\ref{sec:duality} we introduce the $N_C$-dependence into the dispersive
amplitudes following Ref.~\cite{Dai:2017uao}.
The semi-local duality is tested by tuning $N_C$ up to 180. We find that it works
well when $N_C$ is large.
The contributions of each resonance that appears in the amplitudes are also studied.
Finally we give a brief summary in Sec.~\ref{sec:summary}.

\section{Scattering amplitudes and {\boldmath$N_C$} dependence}\label{sec:T}
In Ref.~ \cite{Dai:2017uao}, the $\pi\pi$ scattering amplitudes with $IJ=00,11,02$
(with $I/J$ the total isospin/angular momentum) have already been given. Here, we focus on the
isospin-2 waves with $IJ=20,22$ to complete the analysis. All these waves are certainly
needed for testing the semi-local duality. We use (for more details on the method,
see~\cite{Dai:2017uao}),
\be\label{eq:TP}
T^I_J(s)=P^I_J(s)\Omega^I_J(s)\,,
\ee
with $\Omega^I_{J}(s)$ the Omn\`es function~\cite{Omnes1958}:
\be\label{eq:Omnes}
\Omega^I_{J}(s)=\exp\left(\frac{s}{\pi} \int^\infty_{4M_\pi^2} ds'
\frac{\varphi^I_{J}(s')}{s'(s'-s)}\right) \,.
\ee
Here, $\varphi^I_{J}(s)$ is the phase of the partial wave amplitude
$T^I_{J}(s)$, as given in previous amplitude
analysis~\cite{DLY-MRP14,DLY2015:1}. By a fit to the experimental data
\cite{OPE1973} as well as the amplitudes of the dispersive analysis
\cite{KPY}, the phase is obtained up to $s=4$~GeV$^2$. Above this
energy region we use unitarity to constrain it, but for practical
reasons the extension is limited and we truncate the integration of
the Omn\`es function at $s=22$~GeV$^2$. The other function
$P^I_J(s)$ is represented by a series of polynomials. It absorbs the
contribution from the left-hand cut (l.h.c) and the distant right-hand
cut (r.h.c) above 4~GeV$^2$. To include the Adler zero in the S-wave and
threshold behavior in the D-wave, in terms of the scattering length and
effective range, we parameterize the $P^I_J(s)$ as
\be\label{eq:P}
P^I_J(s)=(s-z^I_J)^{n_J}\sum_{k=1}^n {\alpha^I_J}_k (s-4M_\pi^2)^{k-1}\,,
\ee
with $z^I_J$ to be either the Adler zero
for the S-wave or $4M_\pi^2$ for the D-wave. Similarly, $n_J$ is 1 for the
S-wave and 2 for the D-wave. The fitted parameters $\alpha_i$ are given
in Tab.~\ref{tab:para;alpha}. The units of the $\alpha_k$ are chosen
to guarantee the amplitude $T^I_J(s)$ to be dimensionless.
\begin{table}[htbp]
\hspace{-1.5cm}
\vspace{-0.0cm}
{\footnotesize
\begin{center}
\tabcolsep=0.11cm
\begin{tabular}  {|c|c|c|}
\hline
                                                                & ${T}^2_0(s)$  &  ${T}^2_2(s)$ \rbox   \\[0.5mm] \hline
\rule[0.32cm]{0.1cm}{0cm}$\alpha_1$\rule[0.32cm]{0.1cm}{0cm}    & $-$1.2489       & 0.0472                  \\ \hline
\rule[0.32cm]{0.1cm}{0cm}$\alpha_2$\rule[0.32cm]{0.1cm}{0cm}    & 2.1544          & $-$0.4514               \\ \hline
\rule[0.32cm]{0.1cm}{0cm}$\alpha_3$\rule[0.32cm]{0.1cm}{0cm}    &$-$3.2683(7)     &  1.1773(1)              \\ \hline
\rule[0.32cm]{0.1cm}{0cm}$\alpha_4$\rule[0.32cm]{0.1cm}{0cm}    & 3.2207(3)       & $-$1.5165(1)            \\ \hline
\rule[0.32cm]{0.1cm}{0cm}$\alpha_5$\rule[0.32cm]{0.1cm}{0cm}    &$-$1.8749(1)     &  1.0138(1)              \\ \hline
\rule[0.32cm]{0.1cm}{0cm}$\alpha_6$\rule[0.32cm]{0.1cm}{0cm}    & 0.6212(1)       & $-$0.3587(1)            \\ \hline
\rule[0.32cm]{0.1cm}{0cm}$\alpha_7$\rule[0.32cm]{0.1cm}{0cm}    &$-$0.1077(1)     &  0.0638(1)              \\ \hline
\rule[0.32cm]{0.1cm}{0cm}$\alpha_8$\rule[0.32cm]{0.1cm}{0cm}    & 0.0076(1)       & $-$0.0045(1)            \\ \hline
\end{tabular}
\caption{\label{tab:para;alpha}The fit parameters corresponding to Eq.~(\ref{eq:P}).
The uncertainties are given by MINUIT and $\alpha_{1,2}$ are
fixed by the scattering lengths and slope parameters~\cite{KPY,Ke4}.}
\end{center}
}
\end{table}

\begin{figure}[th]
\includegraphics[width=0.48\textwidth,height=0.3\textheight]{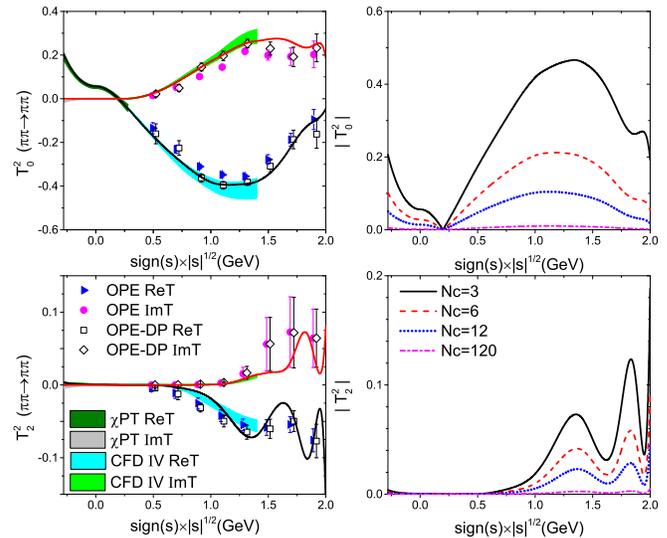}
\caption{\label{Fig:T} Left column: Fit of the isospin-2 $\pi\pi$ scattering amplitudes (solid lines).
  The olive and light grey bands in the low-energy region are from
  $\chi$PT~\cite{Gasser1984,Gasser1985,Bijnens1994,Pelaez02}. The cyan and green bands are from
  CFDIV~\cite{KPY}. The OPE and OPE-DP data are from~\cite{OPE1973}.
  Right column: The absolute values of the amplitudes by varying $N_C$ are shown. The black solid,
  orange dashed, blue dotted and magenta dash-dotted lines are for $N_C=3,6,12,120$, respectively. }
\end{figure}

The fit amplitudes are shown in Fig.~\ref{Fig:T} for the energy
region of $s\in[0,4\,\rm{GeV}^2]$. What we fit to are the following contributions:
$\chi$PT amplitudes for $[0,4M_\pi^2]$
\cite{Gasser1984,Gasser1985,Bijnens1994,Pelaez02}, amplitudes of the
Roy-type equation analysis at $[4M_\pi^2, 2\,\rm{GeV}^2]$
\cite{KPY}, and experiment up to $4\,\rm{GeV}^2$
\cite{OPE1973}. We also plot the amplitudes in the region of $s\in
[-4M_\pi^2,0]$. Here the real part of our amplitudes is in good
agreement with that of $\chi PT$ ($\mathcal{O}(p^4)$), and the
imaginary part vanishes, which is  consistent with the
imaginary part of the $\chi PT$ amplitudes as the latter is rather
small. These indicate the high quality of the fit.

\begin{figure}[!t]
\vspace{-0.0cm}
\includegraphics[width=0.48\textwidth,height=0.3\textheight]{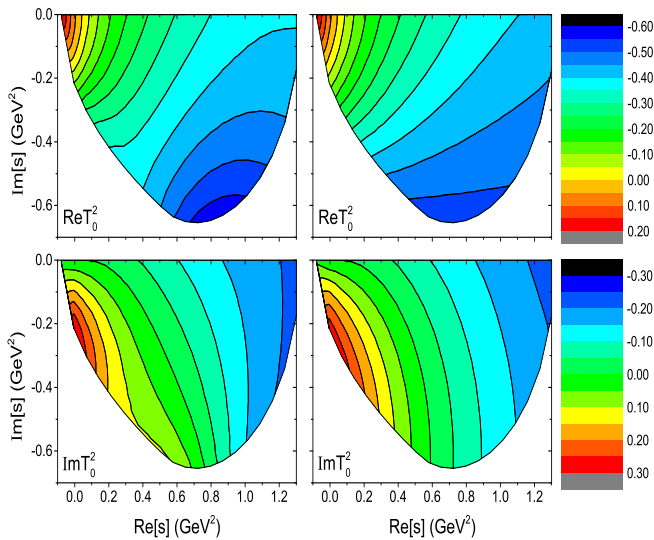}
\caption{Comparison of our amplitudes ($IJ=20$) with the ones
from the Roy-like equation analysis in the domain where the Roy
equations work. On the left side there are real and imaginary parts
of our amplitudes, and on the right side those are from Roy like
equations~\cite{KPY}.
\label{fig:fit;Roy}}
\end{figure}

As is well known, the Roy-type equation analysis embodies  crossing
symmetry\footnote{Notice that the D-wave is absent in the Roy-like
equation analysis~\cite{KPY} and $I=2$ D-wave is very small, we thus
do not discuss it here. For higher partial waves we refer to
Ref.~\cite{Kaminski:2011vj}. }, which is lacking in Eq.~(\ref{eq:TP}).
Therefore, following \cite{Dai:2017tew}, we fit our amplitudes to
the \lq data' on the real axis as well as the amplitudes given by the
Roy-like equation \cite{KPY} in the complex $s$-plane.  As shown in
Fig.~\ref{fig:fit;Roy}, the two $T_0^2$ amplitudes are compatible with each
other except for the region where $s$ is too large (either ${\rm Re}[s]>1.0$~GeV$^2$
or ${\rm Im}[s]<-0.3$~GeV$^2$).  We note that the
amplitudes on the upper half of $s$-plane are readily obtainable
from the ones on the lower side according to the Schwarz reflection
principle. The distribution of contours is in good agreement and
moreover, their gradient variations are compatible with each other,
as shown by the shading of the color from blue to red. Nevertheless,
above 1.0 GeV our amplitudes are a bit different from that of the
dispersive analysis, while both of them are in compatible with the
data, see in Fig.\ref{Fig:T}. Also our amplitudes in the
bottom-right direction, where either $\rm{Re}[s]$ or $\rm{Im}[s]$ is
large, are becoming less consistent with differences $\leq0.1$.

Now that these amplitudes are obtained for the physical world,
that is for $N_C=3$, we can introduce the $N_C$ dependence. Apparently, the real part
of the  $\chi$PT amplitude is $\mathcal{O}(N_C^{-1})$ and the imaginary part is
$\mathcal{O}(N_C^{-2})$ up to any order. Therefore, we generate the
$N_C$ dependence as \cite{Dai:2017uao}:
\be
\varphi(s,N_C) \;=\;\arctan \left[\frac{3}{N_C}\tan \varphi(s) \right] \;, \label{eq:phi;Nc}
\ee
and
\be
P^I_J(s,N_C)\;=\;\frac{3}{N_C}P^I_J(s)\;. \label{eq:P;Nc}
\ee
It is not difficult to check that at large $N_C$ the phase, which would return back
to the phase shift in a single channel,
will jump by $\pi$ around $s=M_R^2$ where $\varphi(M_R^2)=90^\circ$.
This is consistent with the large $N_C$ property of a simple Breit-Wigner formalism or
the ``narrow resonance pole'' on the second Rieman sheet \cite{Guo:2007ff}.
All the complicated higher-order $N_C$ dependence is ignored for
simplicity. By increasing $N_C$, the magnitude of the $I=2$ S-wave and
D-wave become smaller and smaller, as shown in Fig.~\ref{Fig:T}.
This is consistent with the fact that there is no resonance in the
$I=2$ channel.

\section{Semi-local duality}\label{sec:duality}
We follow Ref.~\cite{MRP11} to calculate the variation of the FESR by
tuning $N_C$. It is well-known that the $t$-channel $\pi\pi$
scattering amplitudes could be written in terms of the $s$-channel
ones according to the crossing relations
\bea
T^{I_t}(s,t)=\sum_{I'_s=0}^{2} C_{st}^{I_t I'_s}T^{I'_s}(s,t)\,,
\no \eea
where $I$ denotes the total isospin $(I=0,1,2)$ of the $\pi\pi$-system and $C_{st}$ is the
orthogonal crossing matrix
\bea
C_{st}=C_{ts}= \left(
\begin{array}{c c c}
 \displaystyle 1/3 &  \displaystyle 1 & \displaystyle 5/3 \\
 \displaystyle 1/3 &  \displaystyle 1/2 & \displaystyle-5/6 \\
 \displaystyle 1/3 &  \displaystyle -1/2 & \displaystyle 1/6
\end{array}
\right)\, . \label{eq:Cst}
\eea
The $s$-channel amplitude is composed of a complete
set of partial waves
\bea
T^{I_s}(s,t)=\sum_{J}(2J+1)T_J^{I_s}(s)P_J(\cos\theta_s)\,,\no
\eea
with $\theta_s$ the $s$-channel scattering angle in the center of
mass frame. Of course, $I_s+J$ should be even  as required by
Bose symmetry and isospin conservation. Higher partial waves are
less known and we  restrict our amplitudes up to the D-waves.

We introduce the function
\bea
F^{I_t}_n(\nu,t)={\rm
  Im}T^{I_t}(\nu,t)\nu^{-n}\, ,\label{eq:F;It}
\eea with
$\nu=(s-u)/2$. We note that when $t=4M_\pi^2$, $\nu=s=-u$.
Semi-local duality implies that the contribution of Regge exchange and of
resonances are  dual with each other on the average,
\bea
\int_{\nu_1}^{\nu_2}d\nu F^{I_t}_n(\nu,t)_{\rm resonances}\simeq
\int_{\nu_1}^{\nu_2}d\nu F^{I_t}_n(\nu,t)_{\rm Regge}\,
.\label{eq:dual}
\eea
To test duality in the large $N_C$ limit, we
first estimate it at $N_C=3$. It is helpful to introduce the ratio
\cite{MRP11}
\bea
R^I_n(t)=\frac{\int_{\nu_1}^{\nu_2}d\nu
  F^{I_t}_n(\nu,t)}{\int_{\nu_1}^{\nu_3}d\nu F^{I_t}_n(\nu,t)}\,.
\eea
The upper and lower limits of the integration are chosen as
$\nu_1=(4M_\pi^2+t)/2$, $\nu_2=1$ GeV$^2$, and $\nu_3=2$ GeV$^2$.
The $R^I_n(t)$ of our amplitudes and that of the Regge amplitudes are given in
Tab.~\ref{tab:duality;3}.
\begin{table}[tbp]
{\footnotesize
\begin{tabular}{|c||c|c|c|c|c|}
\hline
\rule[-0.3cm]{0cm}{0.8cm}\multirow{2}{*}{\rule[-1cm]{0cm}{2cm}} &\rule[-0.3cm]{0cm}{0.8cm}\multirow{2}{*}{\rule[-1cm]{0cm}{2cm}~n~}
         &  \multicolumn{2}{c|}{$I_t=0$}   &\multicolumn{2}{c|}{$I_t=1$}  \\
\cline{3-6}
\rule[-0.3cm]{0cm}{0.8cm}    & \rule[-0.3cm]{0cm}{0.8cm}    & $t=4M_\pi^2$ & $t=0$  & $t=4M_\pi^2$ & $t=0$   \\
\hline\hline
\multirow{4}{*}{\rule[-1cm]{0cm}{2.0cm} S,P,D}
     & 0  & $0.431(116)$   & $0.430(122)$  & $0.381(162)$  & $0.396(183)$    \rbox \\[0.5mm]
     & 1  & $0.656(85)$    & $0.668(85)$   & $0.619(131)$  & $0.649(139)$    \rbox \\[0.5mm]
     & 2  & $0.842(40)$    & $0.865(34)$   & $0.829(73)$   & $0.866(69)$    \rbox \\[0.5mm]
     & 3  & $0.948(12)$    & $0.968(8)$    & $0.948(32)$   & $0.973(26)$    \rbox \\[0.5mm]
\hline
\multirow{4}{*}{\rule[-1cm]{0cm}{2.0cm} S,P}
     & 0  & $0.626(201)$   & $0.599(179)$  & $0.779(404)$  & $0.770(384)$    \rbox \\[0.5mm]
     & 1  & $0.801(148)$   & $0.793(130)$  & $0.893(278)$  & $0.896(252)$    \rbox \\[0.5mm]
     & 2  & $0.914(74)$    & $0.921(59)$   & $0.957(132)$  & $0.964(104)$    \rbox \\[0.5mm]
     & 3  & $0.972(26)$    & $0.982(17)$   & $0.987(46)$   & $0.993(29)$     \rbox \\[0.5mm]
\hline
\multirow{4}{*}{\rule[-1cm]{0cm}{2.0cm} Regge}
     & 0  & $0.225$   & $0.233$  & $0.325$   & $0.353$    \rbox \\[0.5mm]
     & 1  & $0.425$   & $0.445$  & $0.578$   & $0.642$    \rbox \\[0.5mm]
     & 2  & $0.705$   & $0.765$  & $0.839$   & $0.908$    \rbox \\[0.5mm]
     & 3  & $0.916$   & $0.958$  & $0.966$   & $0.990$    \rbox \\[0.5mm]
\hline
\hline
\end{tabular}
\caption{\label{tab:duality;3}Comparison of the $R^{I_t}_n(t)$ ratios
between our amplitudes and that of Regge exchange. The latter is given by \cite{MRP11}.
The \lq S,P,D' represent for our work up to the D-waves, and \lq S,P'
with only S- and P-waves. }
}
\end{table}
As can be seen, our calculation with the D-waves is compatible with that of
Regge exchange~\cite{MRP11} within the uncertainties.
For more discussions about the  Regge analysis, we refer to \cite{MRP11}\footnote{It is worth
noting that in \cite{MRP11} the scattering lengths of $IJ=11,02$ waves are calculated in
the Regge parametrization with $n=2,3$. They are in perfect agreement with that obtained by
the dispersive analysis. This certainly confirms the semi-local duality at $N_C=3$, especially
when $n=2,3$. In Ref.\cite{Londergan:2013dza} the non-linear Regge trajectory of the
$f_0(500)$ is obtained and this supports its non-ordinary nature. }.

The difference between the Regge and our amplitudes as well as the difference between our
two results (with or without D-waves)
are much more obvious at $n=0$ than that at $n=3$. This tells us that the D- and even higher
partial waves can not be ignored at small $n$. In contrast,
for large $n$ the low-energy amplitudes will dominate the integration and the contribution
of resonances could be less important. We thus pay attention to $n=1-3$ only and include
the D-waves in next sections. As a support, the $R^I_n(t)$ of the Regge analysis and ours (with D-waves)
are closest to each other for $n=2,3$.
Also we find that the F-waves have tiny contributions only.

\begin{figure}[!tbph]
\vspace{-0.0cm}
\includegraphics[width=0.48\textwidth,height=0.45\textheight]{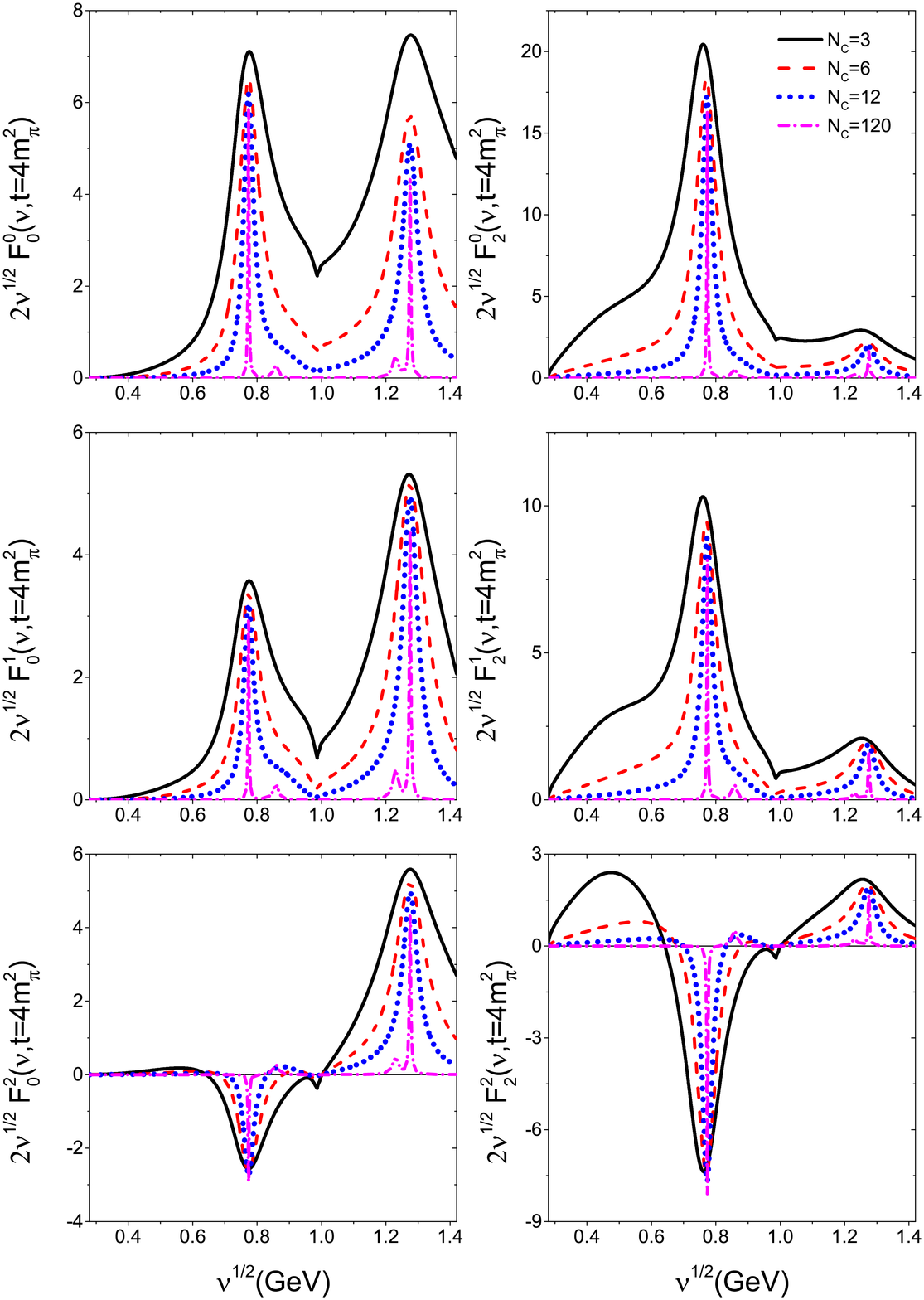}
\caption{The $F^{I_t}_{n}(\nu,t)$ amplitudes for different values of $N_C$, see Eq.~(\ref{eq:Ft}).
  Since $\int d\nu F^{I_{t}}_n(\nu,t)=\int d\sqrt{\nu} ~2\sqrt{\nu}F^{I_{t}}_n(\nu,t)$,
  we plot $2\sqrt{\nu}F^{I_{t}}_n(\nu,t=4M_\pi^2)$ here. \label{Fig:Ft}}
\end{figure}

It is instructive to plot each $F^{I}_{n}(\nu,t)$ amplitude for different values of $N_C$,
see Fig.\ref{Fig:Ft}.
Notice that the peaks around $\sqrt{\nu}=0.85,1.25$~GeV at $N_C=120$ are caused by
the $IJ=00$ wave, cf. Fig.1 of Ref.~\cite{Dai:2017uao}.
They are related to the $f_0(980)$ and the $f_0(1370)$ in the large $N_C$ limit, respectively.
From Fig.~\ref{Fig:Ft} one notices that when $n$ is large and $N_C$ is not too large,
the low energy amplitudes (including the $\sigma$), and the $\rho$ have much larger contributions,
while the $f_2(1270)$ contributes most at $n=0$. Only in the $I_t=2$
amplitude  the contribution of the $\rho$ is negative, which will
cancel other contributions such as from the $\sigma$ in the low energy
region and from the $f_2(1270)$ in the high energy region. This makes sure
that $F^{I_t=2}_n(\nu,t)$ is super-convergent, being much smaller
than the corresponding function for $I_t=0$ or $I_t=1$. Note that resonances/Regge
exchanges are built from $q\bar{q}$ and multi-quark contributions.
When $N_C$ is large, the pole of the $\bar{q}q$ state will fall
down to the real axis on the $s$-plane (zero width), while the
multi-quark component will disappear\footnote{Nowadays it is
believed that  tetra-quarks could be as narrow as the
conventional $\bar{q}q$ resonances \cite{Weinberg2013} or even narrower
\cite{Knecht2013}, however, these won't change our conclusion as we
do not have any tetra-quark in the $I=2$ amplitude.}. Consequently, the
$I_{s/t}=2$ amplitude is super-convergent at large $N_C$ as it does
not contain any resonances or Regge poles. Also the ratio of the $I_t=1$
FESR compared to that of $I_t=0$ should be 2/3 in the large $N_C$
limit. These are analyzed in \cite{MRP11} within U$\chi$PT and we will
check them with our method to generate the $N_C$-dependence in what follows.

\begin{figure}[ht]
\includegraphics[width=0.48\textwidth,height=0.6\textheight]{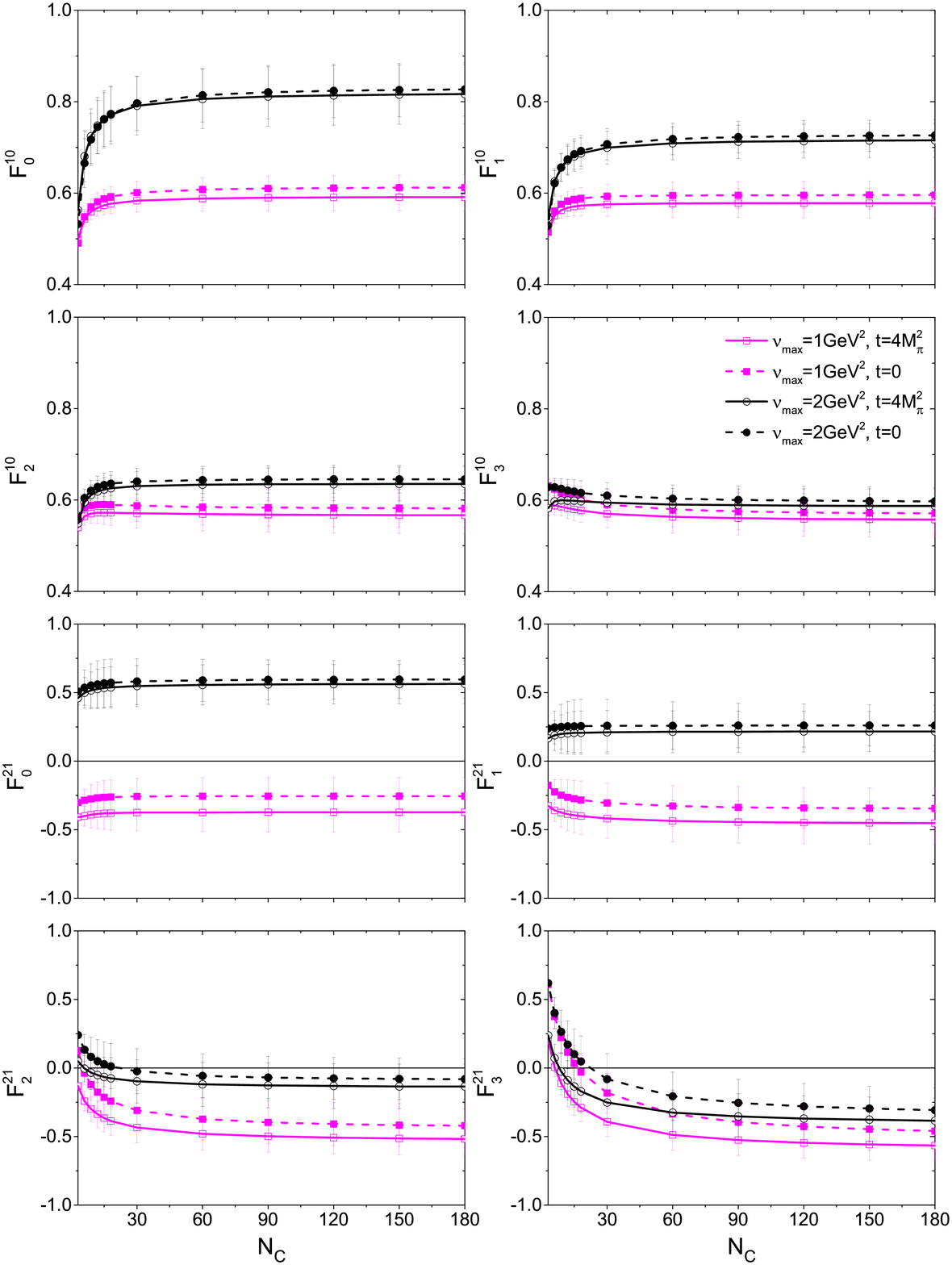}
\caption{\label{Fig:duality;Nc} Duality by varying $N_C$ from 3 to a large number,
  in the order of $N_C=3$, $6$, $9$, $12$, $15$, $18$, $30$, $60$, $90$, $120$, $150$ and $180$.}
\end{figure}

Further, it is helpful to use the definition
\bea
F^{II'}_{n}(t)=\frac{\int_{\nu_{min}}^{\nu_{max}}d\nu F^{I_t}_n(\nu,t)}{\int_{\nu_{min}}^{\nu_{max}}
  d\nu F^{I'_t}_n(\nu,t)}\, .
\label{eq:Ft}
\eea
As discussed before, semi-local duality means that $F^{10}_{n}(t)$ should be  2/3, and
$F^{21,20}_{n}(t)$ be rather small.
The values of these FESR for different values of $N_C$ are given in Tab.~\ref{tab:duality;Nc}.
For convenience, we plot  $F^{10}_{n}(t)$ and  $F^{21}_{n}(t)$ by tuning $N_C$ up to $N_C=180$,
see  Fig.~\ref{Fig:duality;Nc}.

Following \cite{Dai:2017uao}, we simply assume that the whole
contribution of the $N_C^{-2}$ corrections is roughly one third of
that of $N_C^{-1}$, while the correlation between each polynomial is
not discussed, despite the fact that the first two terms of the
polynomials are fixed by the scattering lengths and slope parameters.
In principal, the complete separated $N_C^{-2}$ dependence of each
polynomial in Eq.~(\ref{eq:P;Nc}) could be obtained by matching with
$\chi PT$, if $\chi PT$ is calculated up to higher orders. However,
this is not yet available. One certainly needs a more careful analysis of
the $N_C^{-2}$ corrections\footnote{We note that Ref.~\cite{Jacob2015}
points out that the sub-leading corrections of the low energy constants
(LECs) may be sizable as $L_i/N_C$, which is consistent with our
assumptions. In Refs.~\cite{Pelaez04,Nieves:2009kh}, the uncertainty
caused by the regularization scale $\mu$ is discussed, which is not
required here.}. These higher-order $N_C$ corrections contribute most
to the uncertainties at large $N_C$, estimated by randomly choosing
$2/N_C+3/N_C^2$ and/or $4/N_C-3/N_C^2$ to replace $3/N_C$ in
Eqs.~(\ref{eq:phi;Nc},\ref{eq:P;Nc}) for each partial wave.
The other contributions to the uncertainties are the higher partial
waves,  for instance the $IJ=13$ wave, whose amplitude has been given in
\cite{KPY}, and the systematic uncertainties from different
solutions of the scattering amplitudes. The combination of all the three parts
are collected as the total uncertainties, see  Tab.~\ref{tab:duality;Nc}
for $N_C=3,6,12,120$. The uncertainties are
also shown as error bars in Fig.~\ref{Fig:duality;Nc}. We note that
the uncertainty of the FESR increases as $N_C$ increases.
See e.g. $F^{10,21}_{2}(4M_\pi^2)$ in Tab.~\ref{tab:duality;Nc}, with the upper limit
$\nu_{max}=2$~GeV$^2$. This is because the uncertainty given by
higher-order $N_C$ corrections is important. Besides, the
uncertainty of the $F^{21}_{n}(t)$ is larger than that of
$F^{10}_{n}(t)$. The reason is that the cancellation happens at
$I_t=2$ and they are more sensitive to the relative difference of
each partial waves. The uncertainty coming from  the upper limit
$\nu_{max}=1$~GeV$^2$ is smaller than that of $\nu_{max}=2$~GeV$^2$,
this is caused by the important D-waves. We will discuss this later.

\begin{table}[thbp]
{\footnotesize
\begin{tabular}{|c|c|c|c|c|c|c|}
\hline
\rule[-0.3cm]{0cm}{0.8cm}\multirow{2}{*}{\rule[-1cm]{0cm}{2cm}} &\rule[-0.3cm]{0cm}{0.8cm}\multirow{2}{*}{\rule[-1cm]{0cm}{2cm}~n~}
 &\rule[-0.3cm]{0cm}{0.8cm}\multirow{2}{*}{\rule[-1cm]{0cm}{2cm}~$N_C$~} &  \multicolumn{2}{c|}{$\nu_{max}=1$~GeV$^2$}   &\multicolumn{2}{c|}{$\nu_{max}=2$~GeV$^2$}  \\
\cline{4-7}
\rule[-0.3cm]{0cm}{0.8cm}    & \rule[-0.3cm]{0cm}{0.8cm} & \rule[-0.3cm]{0cm}{0.8cm}   & $t=4M_\pi^2$ & $t=0$  & $t=4M_\pi^2$ & $t=0$   \\
\hline\hline
\multirow{16}{*}{\rule[-1cm]{0cm}{5.0cm} $F_n^{10}$}
     &\multirow{4}{*}{\rule[-1cm]{0cm}{2.0cm}0 }  &3  & $0.50(2)$   & $0.49(2)$  & $0.56(5)$   & $0.53(5)$    \rbox \\
     &                                            &6  & $0.54(2)$   & $0.55(2)$  & $0.68(6)$   & $0.67(5)$    \rbox \\
     &                                            &12 & $0.57(2)$   & $0.58(2)$  & $0.75(6)$   & $0.74(6)$    \rbox \\
     &                                            &120& $0.59(3)$   & $0.61(3)$  & $0.81(6)$   & $0.82(6)$    \rbox \\
\cline{2-7}
     &\multirow{4}{*}{\rule[-1cm]{0cm}{2.0cm}1 }  &3  & $0.51(2)$   & $0.51(2)$  & $0.55(3)$   & $0.53(3)$    \rbox \\
     &                                            &6  & $0.55(2)$   & $0.56(2)$  & $0.63(3)$   & $0.62(3)$    \rbox \\
     &                                            &12 & $0.56(2)$   & $0.58(2)$  & $0.67(3)$   & $0.67(3)$    \rbox \\
     &                                            &120& $0.58(3)$   & $0.60(3)$  & $0.71(4)$   & $0.72(3)$    \rbox \\
\cline{2-7}
     &\multirow{4}{*}{\rule[-1cm]{0cm}{2.0cm}2 }  &3  & $0.54(2)$   & $0.56(2)$  & $0.55(2)$   & $0.55(2)$    \rbox \\
     &                                            &6  & $0.56(2)$   & $0.58(2)$  & $0.60(2)$   & $0.60(2)$    \rbox \\
     &                                            &12 & $0.57(2)$   & $0.59(2)$  & $0.62(2)$   & $0.63(2)$    \rbox \\
     &                                            &120& $0.57(4)$   & $0.58(4)$  & $0.63(4)$   & $0.65(3)$    \rbox \\
\cline{2-7}
     &\multirow{4}{*}{\rule[-1cm]{0cm}{2.0cm}3 }  &3  & $0.58(2)$   & $0.63(2)$  & $0.58(2)$   & $0.63(2)$    \rbox \\
     &                                            &6  & $0.59(2)$   & $0.62(2)$  & $0.60(2)$   & $0.63(2)$    \rbox \\
     &                                            &12 & $0.58(2)$   & $0.61(2)$  & $0.60(2)$   & $0.62(2)$    \rbox \\
     &                                            &120& $0.56(4)$   & $0.57(4)$  & $0.59(4)$   & $0.60(3)$    \rbox \\
\hline\hline
\multirow{16}{*}{\rule[-1cm]{0cm}{5.0cm} $F_n^{21}$}
     &\multirow{4}{*}{\rule[-1cm]{0cm}{2.0cm}0 }  &3  & $-0.41(2)$    & $-0.30(2)$     & $0.46(5)$    & $0.50(5)$   \rbox \\
     &                                            &6  & $-0.40(7)$    & $-0.29(8)$     & $0.50(11)$   & $0.53(13)$   \rbox \\
     &                                            &12 & $-0.39(11)$   & $-0.27(12)$    & $0.53(14)$   & $0.56(17)$   \rbox \\
     &                                            &120& $-0.37(14)$   & $-0.26(14)$    & $0.56(14)$   & $0.59(14)$   \rbox \\
\cline{2-7}
     &\multirow{4}{*}{\rule[-1cm]{0cm}{2.0cm}1 }  &3  & $-0.33(2)$    & $-0.18(2)$     & $0.17(2)$    & $0.24(2)$   \rbox \\
     &                                            &6  & $-0.36(8)$    & $-0.22(9)$     & $0.19(11)$   & $0.25(12)$   \rbox \\
     &                                            &12 & $-0.39(12)$   & $-0.26(13)$    & $0.20(15)$   & $0.25(18)$   \rbox \\
     &                                            &120& $-0.45(15)$   & $-0.34(15)$    & $0.22(15)$   & $0.26(16)$   \rbox \\
\cline{2-7}
     &\multirow{4}{*}{\rule[-1cm]{0cm}{2.0cm}2 }  &3  & $-0.13(2)$     & $0.13(2)$     & $0.05(2)$    & $0.24(2)$   \rbox \\
     &                                            &6  & $-0.24(7)$     & $-0.04(9)$    & $-0.00(10)$  & $0.13(11)$    \rbox \\
     &                                            &12 & $-0.34(10)$    & $-0.18(12)$   & $-0.05(14)$  & $0.05(16)$    \rbox \\
     &                                            &120& $-0.51(12)$    & $-0.41(16)$   & $-0.13(16)$  & $-0.07(15)$   \rbox \\
\cline{2-7}
     &\multirow{4}{*}{\rule[-1cm]{0cm}{2.0cm}3 }  &3  & $0.20(2)$      & $0.61(2)$     & $0.24(2)$    & $0.62(2)$   \rbox \\
     &                                            &6  & $0.01(7)$      & $0.37(9)$     & $0.07(10)$   & $0.40(11)$    \rbox \\
     &                                            &12 & $-0.19(9)$     & $0.11(12)$    & $-0.09(13)$  & $0.17(17)$    \rbox \\
     &                                            &120& $-0.55(11)$    & $-0.43(15)$   & $-0.37(17)$  & $-0.28(17)$    \rbox \\
\hline
\hline
\end{tabular}
\caption{\label{tab:duality;Nc}Duality by tuning $N_C$, given here for $N_C=3$, $6$, $12$, $120$. }
}
\end{table}

The results are quite different when the upper limit is chosen to be
1~GeV$^2$ or 2~GeV$^2$, especially in terms of $n=0,1$. As an
example, for $F^{21}_{0,1}(t)$ the sign of the results with these
two upper limits are even opposite. For $n=2,3$, the differences are
still distinct but smaller. This is consistent with our analysis
that the $IJ=02$ wave contributes a lot in the resonance region at small
$n$ (0 or 1). In Fig.~\ref{Fig:Ft}, the peaks of the $f_2(1270)$ in
$F^{I}_{0}(\nu,t)$ are much larger than those in $F^{I}_{2}(\nu,t)$.
While at large $n$ ($n=2,3$), the contribution of the $IJ=02$ wave in
the resonance region is still important but smaller. Therefore, we
consider the upper limit of 2~GeV$^2$ as the optimal choice.

The change of the results with different $n$ of $F^{10}_{n}(t)$ is
smaller than that of $F^{21}_{n}(t)$ in the large $N_C$ case. For
example, with upper limit $\nu_{max}=2$~GeV$^2$ and $N_C=120$, the
difference between $F^{10}_{0}(4M_\pi^2)$ and  $F^{10}_{3}(4M_\pi^2) $ is 0.22,
while the difference between $F^{21}_{0}(4M_\pi^2)$ and  $F^{21}_{3}(4M_\pi^2)$
is 0.93! The reason is that the contribution from the $\rho$
dominates both amplitudes in $I_t=0$ and $I_t=1$ below 1~GeV, and
the relative sign between different resonances ($\rho$,$f_2(1270)$ etc.) are positive, while
in $I_t=2$ the relative sign is negative, see also Eq.~(\ref{eq:Cst}). This can be checked in
Fig.~\ref{Fig:Ft}, by comparing the lines of
$2\sqrt{\nu}F^{0}_2(\nu,t=4M_\pi^2)$ and $2\sqrt{\nu}F^{1}_2(\nu,t=4M_\pi^2)$.

At large $N_C$, none of the absolute values of $F^{21}_{n}(t)$ is
larger than 0.6, and all $F^{10}_{n}(t)$ are distributed in the
region [0.6,0.8]. For $n=2$, with the upper limit 2~GeV$^2$ and
$N_C=180$, both $F^{10}_{2}(4M_\pi^2)=0.64\pm0.04$ and
$F^{21}_{2}(4M_\pi^2)=-0.14\pm0.16$ are very close to the expected value,
$2/3$ and 0, respectively. Similarly, we have $F^{10}_{2}(0)=0.65\pm0.03$ and
$F^{21}_{2}(0)=-0.08\pm0.15$, even a bit closer.
We note that the two kinds of results, with $t=0$ or $t=4M_\pi^2$, are rather similar to each other.
We thus only discuss the case with $t=4M_\pi^2$ in the next sections.
For $n=1,3$ the situation is not so good, but the values are still fairly close to
the expected values. For $n=1$ we have $F^{10}_{1}(4M_\pi^2)=0.72\pm0.04$ and
$F^{21}_{1}(4M_\pi^2)=0.22\pm0.15$, and for $n=3$ we find $F^{10}_{3}(4M_\pi^2)=0.59\pm0.04$
and $F^{21}_{3}(4M_\pi^2)=-0.38\pm0.17$. The $F^{21}_{n}(0,4M_\pi^2)$ at large $N_C$ indicates
that the $f_2(1270)$ (also the $f_0(980)$ and the $f_0(1370)$) will cancel the contribution
of the $\rho(770)$ most at $n=2$. By increasing/decreasing $n$ the cancellation is
less precise, as the masses of these resonances are different.
By dividing with $\nu^n$ the contribution of the $\rho$ and that of the $f_2(1270)/f_0(980)$ are
mismatched, especially around $\nu=M_R^2$.
As discussed in the earlier sections, $n=1-3$, especially $n=2,3$, are the most valuable
cases to check the semi-local duality, one thus concludes that the
results support that the semi-local duality is fulfilled well up to
$N_C=180$. 
There are some other points that could be interesting.
Almost all the lines in Fig.\ref{Fig:duality;Nc} are increased/decreased a bit strongly from $N_C=3$ to
$N_C=30$. One of the reasons is that the physical amplitudes are not
as simple as that just represented by one or two Breit-Wigner
resonances. Other components, such as multi-quark components and
other background, will also contribute a lot when $N_C$ is not far
away from 3.  
Such variation of the lines of $F^{21}_{n=2,3}(t)$ is more obvious than that of $F^{10}_{n=2,3}(t)$, where the latter is within $1/3$ level. 
This is because the former has a strong cancellation between isospin 0 and 1 waves in the s-channel, as shown in Eqs.~(\ref{eq:Cst},\ref{eq:F;It}). The complex $N_C$ relations of scalars enlarge the variation.
We also notice that the lines are very flat in the
region of $N_C\in[100,180]$. It is thus natural to infer that they will
stay flat for larger $N_C$. This suggests that the semi-local
duality will hold in the large $N_C$ limit.

\begin{figure}[ht]
\includegraphics[width=0.48\textwidth,height=0.25\textheight]{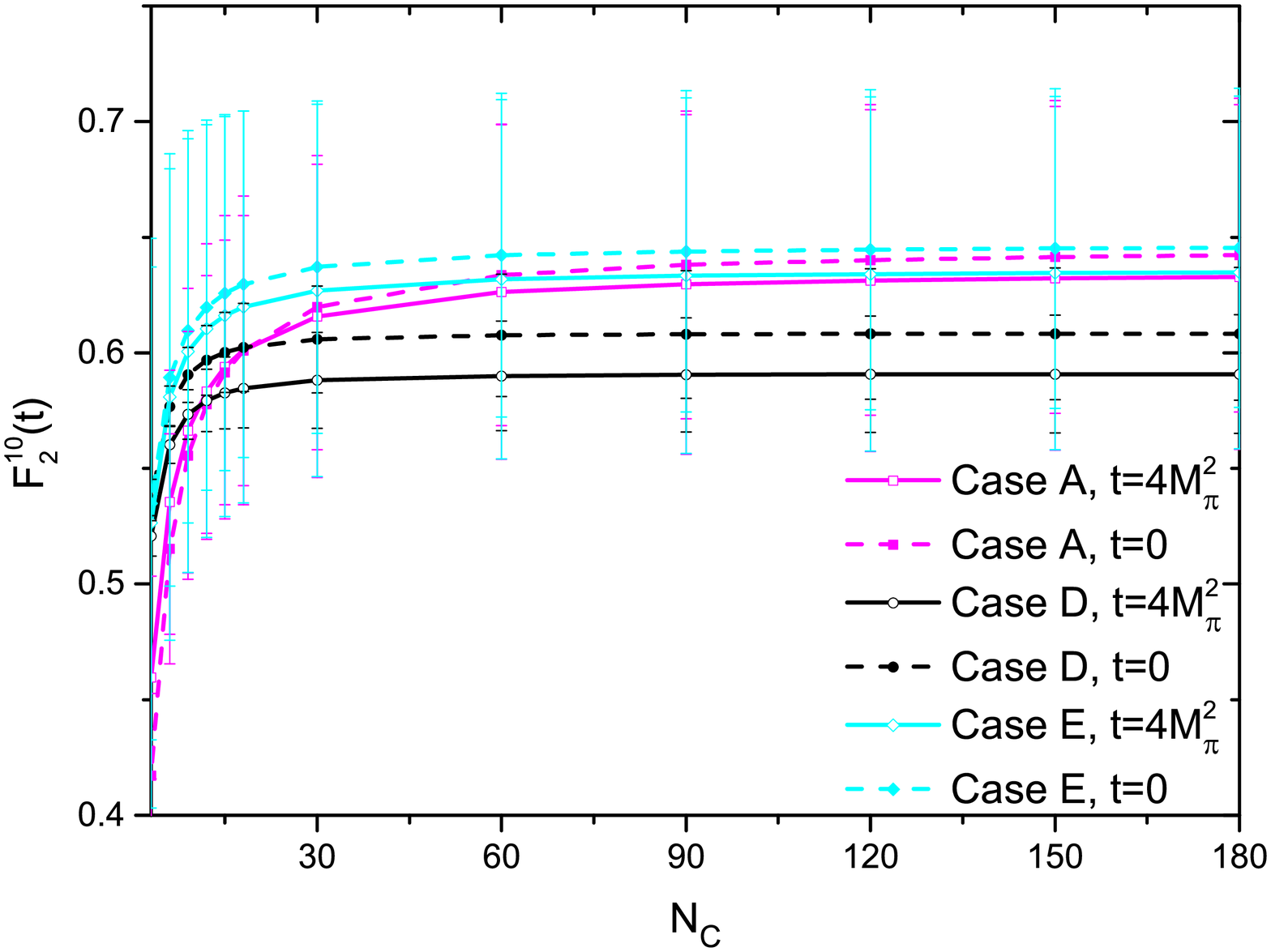}
\includegraphics[width=0.48\textwidth,height=0.25\textheight]{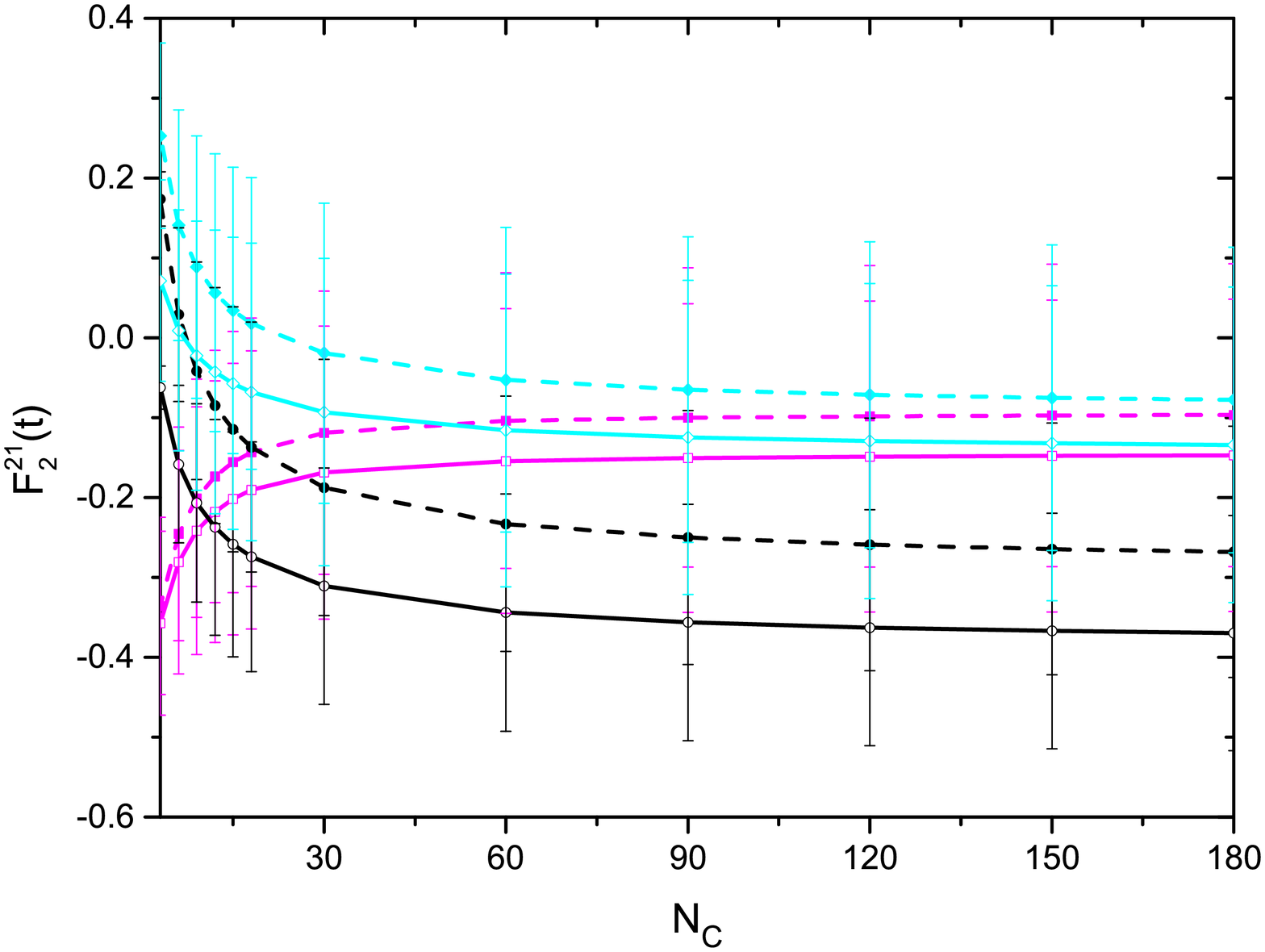}
\caption{\label{Fig:duality;case} Duality for the different cases defined in the text.
  We vary $N_C$ from 3 to a large number, in the order of $N_C=3$, $6$, $9$, $12$, $15$, $18$,
  $30$, $60$, $90$, $120$, $150$ and $180$. }
\end{figure}

To estimate the contribution of each resonance at large $N_C$, we perform the following
calculations:
\begin{itemize}
\item[$\bullet$]{\bf Case~A}: The amplitudes of the $IJ=00$ wave in the region of $\sqrt{s}\leq0.75$~GeV
  have been set to  zero. In this case the main contribution of $f_0(500)$ is removed.
\item[$\bullet$]{\bf Case~B}: Similar to Case~A, the amplitudes of the $IJ=00$ wave in the region
  of $\sqrt{s}\in[0.81,0.91]$~GeV have been set to  zero. The $f_0(980)$ is removed  in
  the large $N_C$ limit.
\item[$\bullet$]{\bf Case~C}: Similar to Case~A, the amplitudes of the $IJ=00$ wave in the region
  of $\sqrt{s}\in[1.13,1.33]$~GeV have been set to zero. The $f_0(1370)$ is removed in the large
  $N_C$ limit.
\item[$\bullet$]{\bf Case~D}: Similar to Case~A, the amplitudes of the $IJ=02$ wave in the region
  of $\sqrt{s}\in[1.15,1.4]$~GeV have been set to zero. The contribution of the $f_2(1270)$ is removed.
\item[$\bullet$]{\bf Case~E}: Only the upper limit is changed to $\nu_{max}=4$~GeV$^2$, where the
  possible contribution of heavier resonances, like $\rho(1450)$, $\rho(1710)$ etc. is included.
\end{itemize}
The results are shown in Tab.~\ref{tab:duality;case}.
For Cases A, D and E, the results by tuning $N_C$ are plotted as magenta, black, and cyan lines,
respectively, in Fig.~\ref{Fig:duality;case}.
For Cases B and C, we can not extract  the contributions of the relevant resonances
except at large $N_C$, as there is no obvious peak for $f_0(980)$ and $f_0(1370)$ around $N_C=3$.
Therefore, we only show the results at $N_C=180$, see Tab.~\ref{tab:duality;case}.
\begin{table}[htbp]
{\footnotesize
\begin{tabular}{|c|c|c|c|c|}
\hline
\rule[-0.3cm]{0cm}{0.8cm}\multirow{2}{*}{\rule[-1cm]{0cm}{2cm}~Case~} &  \multicolumn{2}{c|}{$F^{10}_2(t)$}   &\multicolumn{2}{c|}{$F^{21}_2(t)$}  \\
\cline{2-5}
  \rule[-0.3cm]{0cm}{0.8cm}   & $t=4M_\pi^2$ & $t=0$  & $t=4M_\pi^2$ & $t=0$   \\
\hline\hline
{\bf O}  & $0.64(5)$   & $0.65(4)$  & $-0.14(16)$   & $-0.06(15)$    \rbox \\
\hline
     A   & $0.63(7)$   & $0.64(7)$  & $-0.15(20)$   & $-0.10(19)$    \rbox \\
\hline
     B   & $0.59(8)$   & $0.59(7)$  & $-0.36(20)$   & $-0.34(19)$    \rbox \\
\hline
     C   & $0.62(8)$   & $0.63(7)$  & $-0.23(20)$   & $-0.19(19)$    \rbox \\
\hline
     D   & $0.59(3)$   & $0.61(3)$  & $-0.37(15)$   & $-0.27(16)$    \rbox \\
\hline
     E   & $0.63(8)$   & $0.65(7)$  & $-0.13(20)$   & $-0.08(19)$    \rbox \\
\hline
\hline
\end{tabular}
\caption{\label{tab:duality;case}Duality for the different Cases at $N_C=180$ discussed in the text.
  Note that Case O is the original result shown in Fig.~\ref{Fig:duality;Nc}.  }
}
\end{table}

For Case A, we have $F^{10}_2(4M_\pi^2)=0.63\pm0.07$ and $F^{21}_2(4M_\pi^2)=-0.15\pm0.20$ at $N_C=180$.
These satisfy the semi-local duality well. At $N_C=3$, we have $F^{21}_2(4M_\pi^2)=-0.36\pm0.12$.
It changes a lot comparing to the original result, cf. Tab.~\ref{tab:duality;Nc}. It confirms
that the $\sigma$ contribution is not ignorable at $N_C=3$, while it is smaller and irrelevant
in the large $N_C$ limit. The cancellation does not happen between the $\sigma$ and $\rho(770)$ in
the large $N_C$ limit.
This also suggests that $\sigma$ is dominated by the non-$\bar{q}q$ structure components.

It is worth to point out that in Ref.~\cite{Guo:2012yt}, where U$\chi$PT based on the
N/D method has been applied to the meson-meson scattering of the $U(3)$ nonet,
the $f_0(980)$ contribution can not be ignored to cancel the contribution of $\rho$.
In contrast, in Ref.\cite{MRP11}, where U$\chi$PT is realized by inverse amplitude method,
the $f_0(980)$ is irrelevant and a sub-dominant $\bar{q}q$ component is needed for the cancellation.
For the $f_0(1370)$, both of these two
works agree as the resonance does not contribute a lot at large $N_C$.
In our case,  the two resonances behave as \lq peaks' around 0.85~GeV
and 1.23~GeV at large $N_C$, respectively. This supports their possible inner
$\bar{q}q$ component, resulting in a possible contribution to cancel
the $\rho$ in $F^{21}_{n}(t)$. We find numerically for  Case~B, where
the contribution of the $f_0(980)$ has been removed  at large
$N_C$, that $F^{10}_2(4M_\pi^2)=0.59\pm0.08$ and
$F^{21}_2(4M_\pi^2)=-0.36\pm0.20$ at $N_C=180$. Comparing to the
original results (Case~{\bf O}), the value of $F^{10}_2(0,4M_\pi^2)$
change a bit, and that of $F^{21}_2(0,4M_\pi^2)$ change a lot. Thus, the
contribution of the $f_0(980)$ at large $N_C$ can  not be ignored. For Case~C,
by removing the $f_0(1370)$ at large $N_C$, the change is smaller
than that of Case B but still distinct, see  Tab.~\ref{tab:duality;case}.
This implies that the $f_0(980)$ and the  $f_0(1370)$
have a significant $\bar{q}q$ component and they will partly cancel the
contribution of the $\rho$.

In Case~D the $f_2(1270)$ has been removed. One finds $F^{10}_2(4M_\pi^2)=0.59\pm0.03$ and
$F^{21}_2(4M_\pi^2)=-0.37\pm0.15$ at $N_C=180$. The $f_2(1270)$ contribution is rather
important to cancel that of the $\rho$, just as expected. The results of Case~D is rather close
to that of the $f_0(980)$ at large $N_C$,
implying the same important contribution of the $f_0(980)$ as the $f_2(1270)$.
In case~E we consider the FESR with the upper limit $\nu_{max}=4$~GeV$^2$.
The results at $N_C=180$ are almost the same as that of Case~{\bf O}.
This supports the view that the $\rho(1450)$ as well as other heavier resonances do not have
a large effect. In Fig.~\ref{Fig:duality;case} one clearly sees that in Case~D the $F^{10}_2(t)$
and $F^{10}_2(t)$ deviate much more from the expected values than
that in Cases~A and E. This confirms  that the $f_2(1270)$ has a much larger effect than the
$f_0(500)$ and heavier resonances such as the $\rho(1450)$.

\section{Summary}\label{sec:summary}
In this paper we have studied the semi-local duality for large $N_C$.
The isospin-2 $\pi\pi$ scattering amplitudes with final-state
interactions are constructed in a model-independent way and fit to
the data. Comparing with the amplitudes of $\chi$PT, we generate the $N_C$-dependence
of the amplitudes. With this $N_C$-dependence
the semi-local duality in terms of finite energy sum rules is
tested. Our results show that the semi-local duality is satisfied
well in the large $N_C$ limit, at least up to $N_C=180$. This study
confirms that the method  of generating the $N_C$ dependence proposed
in Ref.~\cite{Dai:2017uao} is reliable. At large $N_C$, the contributions of
the $f_2(1270)$ and the $f_0(980)$ are important to cancel
that of the $\rho(770)$, the latter is consistent with what has been found in Ref.~\cite{Guo:2012yt}. Also the $f_0(1370)$  contributes significantly to the cancellation.
In contrast, the $f_0(500)$ (or $\sigma$) has a
large effect for $N_C=3$ and a small effect at large $N_C$. These support the $\bar{q}q$ component
in the  $f_0(980)$ and the $f_0(1370)$, but not for the  $\sigma$ as required in Ref.~\cite{MRP11}.

\section*{Acknowledgements}
We are very grateful to Prof. Michael R. Pennington, who has just passed away.
Through discussions with him the idea underlying this paper was generated.
Helpful discussions with Profs. Han-Qing Zheng and Zhi-Hui Guo are also acknowledged.
This work is supported by the DFG (SFB/TR 110, ``Symmetries and the Emergence of
Structure in QCD''), by the Chinese Academy of Sciences (CAS) President's
International Fellowship Initiative (PIFI) (Grant No. 2018DM0034) and the
VolkswagenStiftung (Grant No. 93562).


\end{document}